# Typical Physics PhD Admissions Criteria Limit Access to Underrepresented Groups but Fail to Predict Doctoral Completion

Casey W. Miller, [1*] Benjamin M. Zwickl, [2] Julie R. Posselt, [3]
Rachel T. Silvestrini, [4] Theodore Hodapp [5]

[1]School of Chemistry and Materials Science, Rochester Institute of Technology
85 Lomb Memorial Drive, Rochester, NY, 14623, USA
[2]School of Physics and Astronomy, Rochester Institute of Technology
85 Lomb Memorial Drive, Rochester, NY, 14623, USA
[3]Rossier School of Education, University of Southern California,
3470 Trousdale Parkway, Los Angeles, CA 90089, USA
[4]Industrial and Systems Engineering Department, Rochester Institute of Technology
85 Lomb Memorial Drive, Rochester, NY, 14623, USA
[5]American Physical Society, One Physics Ellipse, College Park, MD 20740, USA

[*]To whom correspondence should be addressed; E-mail: cwmsch@rit.edu

This work aims to understand how effective the typical admissions criteria used in physics are at identifying students who will complete the PhD. Through a multivariate statistical analysis of a sample that includes roughly one in eight students who entered physics PhD programs from 2000-2010, we find that the traditional admissions metrics of undergraduate GPA and the Graduate Records Examination (GRE) Quantitative, Verbal, and Physics Subject Tests do not predict completion in US physics graduate programs with the efficacy often assumed by admissions committees. We find only undergraduate GPA to have a statistically significant association with physics PhD completion across all models studied. In no model did GRE Physics or GRE Verbal predict PhD completion. GRE Quantitative scores had statistically significant relationships with PhD completion in two of four models studied. However, in practice, probability of completing the PhD changed by less than 10 percentage points for students scoring in the $10^{th}$ vs $90^{th}$ percentile of US test takers that were physics majors. Noting the significant race, gender, and citizenship gaps in GRE scores, these findings indicate that the heavy reliance on these test scores within typical PhD admissions process is a deterrent to increasing access, diversity, and equity in physics. Misuse of GRE scores selects against already-underrepresented groups and US citizens with tools that fail to meaningfully predict PhD completion.    This is a draft; see the journal for the published version.

**Additionally included in blue text are several responses to queries about this work.**

# Introduction

Physics is the least diverse of the sciences, rivaling mechanical engineering and aerospace engineering for the least diverse fields within all of science, technology, mathematics, and engineering (STEM) [1]. Groups underrepresented in physics include Blacks, Latinos, Native Americans, and women of all racial/ethnic groups. Barely 5% of physics PhDs are granted annually to those identifying with an underrepresented racial/ethnic category; women earn only 20% of physics PhDs. The origins of these vast representation gaps are complex and include inequitable educational access from an early age [2], implicit bias in the classroom and research laboratories [3], deterrents to continuation for underrepresented groups (e.g., departmental climate and disciplinary culture) [4, 5, 6], and stereotype threat [7, 8]. Expanding gender and racial participation in STEM is important for the development of a robust domestic scientific workforce, however, as pointed out by the National Academy of Sciences report *Expanding Underrepresented Minority Participation: America's Science and Technology Talent at the Crossroads* [9]. Who gets to do the science of the future is largely determined by who is selected into PhD programs. The transition to graduate programs is thus a concern of national importance; only by attending to structural issues present in the process of selecting who gets to do the science of the future can we make sustainable progress toward broadening the participation of groups historically underrepresented in STEM.

Unfortunately, non-trivial barriers impede admission to PhD programs for some demographic groups. Undergraduate grades, college selectivity, and GRE scores are the three criteria that best predict admission to US graduate programs [10], but are not evenly distributed by race and gender [10, 11]. This is particularly problematic for easily sortable numeric metrics, such as GRE scores. Predictive validity analyses of the GRE are almost as old as the test itself [12, 13, 14]. Research over decades of test refinement, as well as meta-analysis of such research, consistently finds that scores on the Verbal and Quantitative GRE (GRE-V; GRE-Q) have weaker validity for PhD attainment than for graduate school grades [15]. Using the same database as Ref. kuncel2001comprehensive, additional analysis identified positive relationships between these tests' scores and first-year grades, cumulative grades, and faculty ratings [16]. In a similar vein, two recent studies on biomedical PhD admissions found that the General GRE does not predict scholarly productivity [17] or degree completion, but that scores are associated with first semester and cumulative graduate school grades [18]. Methodologically, most assessments of validity focus on the general test and are limited to bivariate correlation analyses; they do not include covariates to render more precise estimates. Overall, the record indicates that the GRE's validity wanes as time elapses between taking the test and measuring "success" in graduate school, which may be indicated by completion, research productivity, and other markers of success.

Despite their near universal employment by physics PhD programs [19], no study has tested the validity of common admissions metrics explicitly in these programs. Given the strong race, gender, and citizenship performance differences on the GREs in particular [10, 11], it is critical that we know the extent to which scores are useful in identifying students who will complete the PhD. We conducted such a study, inviting physics programs to submit de-identified student admission and degree completion records. Among applicants that were admitted and matriculated into physics PhD programs, we find the predictive validity to be poor for some of the most ubiquitously used admissions criteria. In particular, we find undergraduate GPA to be the most robust numerical predictor of PhD completion, and, despite a large sample size and wide dynamic range, we do not find a statistically significant relationship between GRE Physics (GRE-P) Subject Test scores and PhD completion.

This article is structured as follows. First, we provide a snapshot of the state of US physics with respect to diversity and degree production. Next, we describe US citizens' performance on the GRE-P, across a variety of demographic parameters. We then describe our multivariate regression analysis and its findings. We conclude with implications of these results.

Table 1: Multi-year averages for diversity metrics in US physics. Each row's entries show the percentage of the US annual average in the second column. Data for the GRE-P are the average over test years 2009-2015; all others are the averages from 2009-2014 (20). We excluded data if an entry's absolute number was ten or less (indicated by *), and excluded race categories "other", "two or more", and "no response". F = female; M = male.

|  | US Annual Average | US | | Hispanic | | Asian | | Black | | White | | Native Am. | | non-US | |
| --- | --- | --- | --- | --- | --- | --- | --- | --- | --- | --- | --- | --- | --- | --- | --- |
|  |  | F | M | F | M | F | M | F | M | F | M | F | M | F | M |
| Bacc. Degrees | 5837 | 19% | 81% | 1.2% | 5.0% | 1.6% | 5.5% | 0.7% | 2.1% | 14% | 61% | * | 0.4% | - | - |
| GRE-P Test Takers | 2914 | 20% | 80% | 1.2% | 4.3% | 1.9% | 6.3% | 0.4% | 1.3% | 16% | 63% | * | 0.4% | 570 | 2073 |
| PhD Matriculants | 1550 | 18% | 82% | 1.0% | 4.8% | 1.4% | 5.2% | * | 1.6% | 14% | 63% | * | * | 266 | 897 |
| PhDs Awarded | 960 | 18% | 82% | * | 3.5% | 2.0% | 5.5% | * | 1.6% | 13% | 61% | * | * | 182 | 681 |
| GRE-P Test Takers per Bacc. Degree | 50% | 52% | 49% | 51% | 43% | 57% | 57% | 26% | 30% | 55% | 51% | * | 42% | - | - |
| PhD Matriculants per Bacc. Degree | 27% | 25% | 27% | 23% | 25% | 23% | 25% | * | 20% | 26% | 27% | * | * | - | - |
| PhD Completion Rate | 62% | 62% | 62% | * | 46% | 86% | 65% | * | 60% | 59% | 60% | * | * | 68% | 76% |

## Current State of US Physics

The state of diversity in physics can be summarized by the annual average numbers of bachelor's degrees awarded, first year graduate students, PhDs awarded, and the performance of students on the GRE-P. Whereas the latter data are obtained from ETS itself, the remainder are available through the Integrated Postsecondary Education Data System (IPEDS) [20] . From the IPEDS data (Tab. 0), several observations are possible. At all stages of physics education, Latinos and Blacks are underrepresented relative to their college-age representation in the US, whereas Asians and Whites are overrepresented. The ratio of GRE-P test takers to physics undergraduate degrees awarded indicates that approximately half of physics undergraduate degree earners are actively considering physics graduate studies. About one quarter of US physics majors matriculate into US physics graduate programs. Significant exceptions to these trends are noted for Blacks, who take the GRE-P and matriculate at lower rates than the national average. Black females, Latinas, and Native Americans of any gender each had fewer than ten physics PhD matriculants annually in these data. Finally, women are barely 20% of physics students, at both the undergraduate and graduate levels, and they take the GRE-P in proportion to their representation.

IPEDS data indicate that around 60% of US citizens that matriculate to PhD programs will complete their degree. We do not take into consideration the time dependence of matriculants and PhDs earned, leading to some ambiguity in completion rate. However, we note that the Council of Graduate Schools indicates the 10-year completion rate for physics overall is 59%, close to what we report here [21]. There is no overall gender gap for physics PhD completion or time-to-PhD (not shown) among US citizen graduate students. With the caveat that low enrollment numbers imply a relative error on the of order of 20%, these data indicate that Hispanic males have a lower PhD completion rate than the average, and Asian females have the highest PhD completion rate of US citizens (Tab. 0).

Figure 0 shows significant gaps in GRE-P scores for US citizens based on race and gender. These data, obtained from the ETS database (portal.ets.org), represent all test-takers who earned a valid GRE-P score in test years 2009-2015. The median US female score is a 580 ($28^{th}$ percentile), while the median US male score is a 650 ($46^{th}$ percentile); ETS reports [22] the standard error of measurement to be 49 points (roughly 9 percentile), indicating that gender gaps are statistically significant. Notably, similarly large gender gaps in GRE-P scores exist for all racial/ethnic groups for both US and international test takers (median percentile by country for (male, female) test takers were: China ($86^{th}$, $77^{th}$); India ($70^{th}$, $46^{th}$); Iran ($62^{nd}$, $42^{nd}$)). The median scores for Black (530; $17^{th}$ percentile), Hispanic (580; $28^{th}$ percentile), White (630; $39^{th}$ percentile), and Asian Americans (690; $53^{rd}$ percentile) also reveal significant variation in GRE-P by race.

Although the best evidence suggests that faculty are well intentioned when selecting students, many are unaware of demographic patterns in GRE scores, and they carry out admissions according to inherited practices that include using cut-off scores [23]. Programs using the GRE-P as an integral part of their admissions process may be unwittingly selecting against underrepresented groups and US citizens. This is easily inferred when combining the race, gender, and citizenship score differences with the use of strict cut-offs (or even preferences) based on GRE scores. Unfortunately, this is common practice throughout the disciplines [23] and in physics specifically [19, 24]. Approximately 25 percent of physics PhD programs publicize to potential applicants a minimum acceptable GRE-P score around 700 ($55^{th}$ percentile). The representation of the US test takers above this level is very different from the applicant pool: Hispanics and Blacks are 6.2% and 1.8% of test takers, but only 4.1% and 0.6% of those whose scores exceed 700; Asians are 7.8% of test takers, but their above-700 representation is 11.4%; the representation of Whites is unchanged at 78%; women are 20% percent of test takers, but only 11% of those scoring above 700.

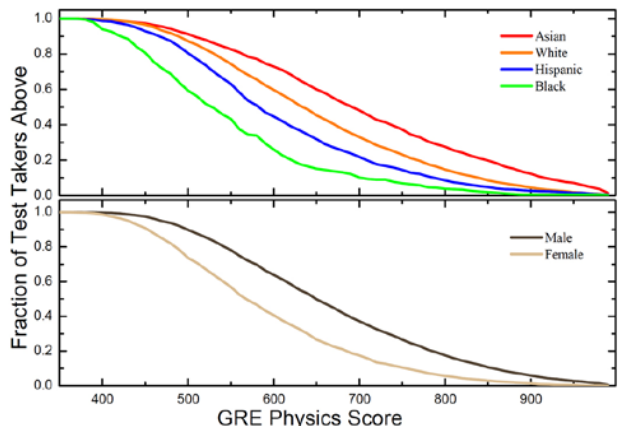

Figure 1: The fraction of US test takers above a specified GRE Physics Score shows that cut-

off scores adversely impact underrepresented groups more than majority groups. Given that we find that PhD completion is not correlated with the Physics GRE score, the misuse of the test in admissions will negatively affect diversity without being able to identify individuals able to complete a physics PhD. Source: ETS.

## Physics PhD Completion Study

The goal of this study was to ascertain which of the common quantitative admissions parameters in physics are significantly correlated with PhD completion. To this end, we requested student level data from all departments that awarded more than 10 PhDs/year. We requested the following information about students who matriculated in 2000 through 2010: undergraduate GPA (UGPA); GRE-Q; GRE-V; GRE-P; graduate GPA (GGPA); final disposition of student (i.e., PhD earned or not); start and finish year; and demographic information. These data were then analyzed with multivariate logistic regression techniques to identify the extent to which the independent parameters can be used to predict PhD completion probability.

We received data from 27 programs (a response rate of about 42%), representing programs with a broad range of National Research Council (NRC) rankings; the data set includes peers and aspirational peers for every type of physics PhD program in the US. We include the doctoral programs' NRC ranking [25] as a categorical variable. As the NRC only gives confidence intervals for program rank, we created a ranking for this study by averaging the 5% and 95% confidence bounds for the NRC regression-based ranking (NRC-R), and rounded this up to the nearest five to protect the confidentiality of participating programs. This led to a ranking range of 5 to 105. We divided the programs into terciles of approximately equal number of records, categorized as Tier 1 (highest ranked, NRC-R≤20), Tier 2 (25≤NRC-R≤55, and Tier 3 (NRC-R>55). By analyzing data from PhD programs whose NRC ranking varies widely, we have data from highly selective to much less selective programs, providing us with greater variation in GRE scores than predictive validity analyses deriving from a single program.

Our analytic sample included all students in 24 programs for which start year was available. We identified start year as a sample inclusion criterion because, it would be impossible to determine whether PhD non-completion was simply due to too few years of enrollment without this. These data cover 3962 students, which corresponds to roughly 13% of matriculants to all US physics PhD programs during the years studied. In the analytic sample, 18.5% are women of any race or citizenship and 58.4% are US citizens. The racial composition of US citizens in the dataset is 63.6% White, 1.3% Black, 2.4% Hispanic, 0.2% Native American, 4.1% Asian, Multiple or other races 0.9% and 27.6% Race Unavailable. Excluding the cases for which race was unavailable, the sample is thus roughly representative of annual PhD production in US physics for gender, race, and citizenship, as indicated in Tab. 0.

We model PhD completion as a function of UGPA (on a four-point scale), GRE scores (GRE-Q, GRE-V, GRE-P), gender (man or woman), citizenship (US or non-US), race/ethnicity, and NRC ranking. We employ multivariate logistic regression to improve upon the correlation coefficient as a measure of validity. Given that multiple parameters may be associated with completion, a multivariate approach allows us to isolate how individual parameters relate to this outcome, controlling for others that may or may not relate. In this case, the logistic regression provides a best fit probability of PhD completion ($p$) vs. non-completion ($1-p$) as a function of independent model parameters. The "$logit$", $l$, associated with each independent variable is defined as the natural log of the odds ratio. As an example, for a univariate model based on UGPA,

$$l(UGPA) = \ln \frac{p(UGPA)}{1-p(UGPA)}.$$

The $logit$ is assumed to vary linearly with the parameter, e.g., $l(UGPA) = a + bUGPA$, where $a$ and $b$ are fit parameters. A model with one independent parameter linking UGPA to completion probability can then be written as $p(UGPA) = 1/(1 + \exp(-a - bUGPA))$. The coefficient associated with UGPA is interpreted as the change in the log of the odds of PhD completion that is associated with a one point increase in GPA. The interpretation is similar in multivariate analyses, but each coefficient estimate also takes into account (i.e., controls for) simultaneous relationships that other model parameters may have with PhD completion. As such, multivariate models provide a more complete, precise picture than bivariate correlation coefficients of what explains an outcome and of individual parameters' relationships with the outcome. It is important to note that bivariate correlation coefficients, while easier to compute, include the influence of confounding factors that are also related to the outcome of interest, and therefore are inadvisable as a basis for policy decisions.

Finally, we use a standard $p$ value of 0.050 or less to gauge statistical significance in this article, recognizing the limitations of this metric [26].

## Results

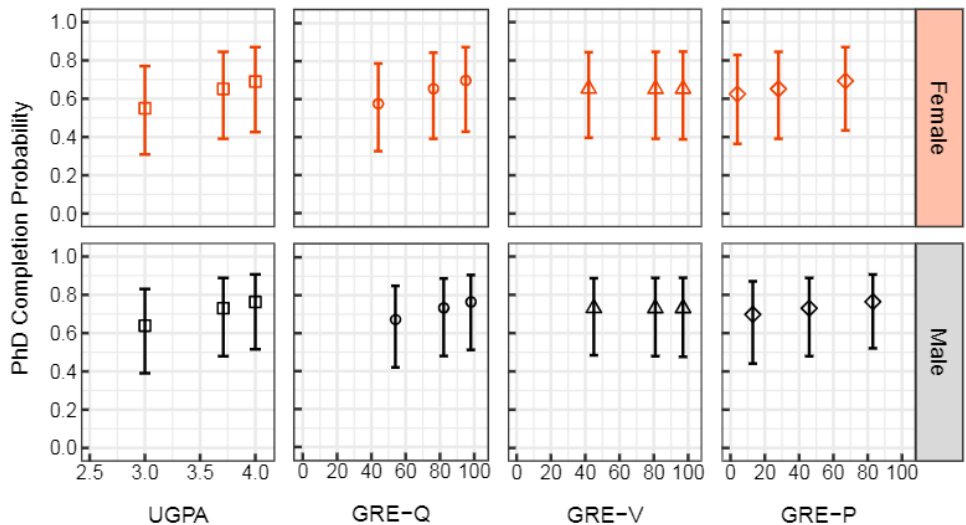

Figure 2: Multivariate logistic regression results for the *US Only* model for UGPA, GRE-Q, GRE-V, and GRE-P for women and men, controlling all other variables (continuous variables held at median values; categorical variables constant). Model results are indicated for the 10[th], 50[th], and 90[th] percentile scores for US women and men test takers whose self-reported intended graduate major was physics or astronomy; this range represents the vast majority of scores that can be anticipated by physics and astronomy graduate admissions committees. The whiskers on the model results indicate the 95% confidence intervals associated with PhD completion probability. The relatively flat model results highlight the subtlety of any relationship.

Table 2: Multivariate logistic regression results modeling physics PhD completion in four analytic samples. The coefficients for logit and odds ratios (OR) for PhD Completion are reported, along with their standard errors (SE). Tier 1 (highest ranked, NRC-R≤20), Tier 2 (25≤NRC-R≤55, and Tier 3 (NRC-R>55). Reference groups are Tier 3 for ranking and White for Race/Ethnicity. * = $p < 0.05$; ** = $p < 0.01$; *** = $p < 0.001$.

|  | All Students (N = 3962) | | US only (N = 2315) | | US Female (N = 402) | | US Male (N = 1913) | |
|---|---|---|---|---|---|---|---|---|
|  | Logit (SE) | OR (SE) | Logit (SE) | OR (SE) | Logit (SE) | OR (SE) | Logit (SE) | OR (SE) |
| (Intercept) | -1.63** (0.53) | 0.2** (0.1) | -2.63*** (0.69) | 0.07*** (0.05) | -4.46** (1.65) | 1E-2** (0.02) | -2.05** (0.77) | 0.1** (0.1) |
| UGPA | 0.31* (0.12) | 1.4* (0.2) | 0.60*** (0.16) | 1.8*** (0.3) | 0.9* (0.4) | 2.5* (1) | 0.47* (0.18) | 1.6* (0.3) |
| GRE-Q | 13E-3** (4E-3) | 1.013** (0.004) | 10E-3* (5E-3) | 1.011* (0.005) | 0.017 (0.012) | 1.02 (0.01) | 0.01 (6E-3) | 1.010 (0.006) |
| GRE-V | -1E-3 (2E-3) | 0.999 (0.002) | -1E-4 (3E-3) | 0.9999 (0.003) | -1E-3 (7E-3) | 0.999 (7E-3) | -5E-06 (3E-3) | 1.000 (0.003) |
| GRE-P | 3E-3 (2E-3) | 1.004 (0.002) | 5E-3 (3E-3) | 1.005 (0.003) | 2E-4 (6E-3) | 1.000 (0.006) | 5E-3 (3E-3) | 1.005 (0.003) |
| Tier 1 | 0.69*** (0.1) | 2.0*** (0.2) | 0.73*** (0.14) | 2.1*** (0.3) | 0.90** (0.34) | 2.5** (0.8) | 0.74*** (0.15) | 2.1*** (0.3) |
| Tier 2 | 0.23* (0.1) | 1.3* (0.1) | 0.53*** (0.13) | 1.7*** (0.2) | 0.15 (0.3) | 1.2 (0.4) | 0.63*** (0.15) | 1.9*** (0.3) |
| Asian | -0.02 (0.28) | 1.0 (0.3) | -0.01 (0.28) | 0.99 (0.28) | 0.09 (0.51) | 1.1 (0.6) | -0.07 (0.34) | 0.9 (0.3) |
| Black | -0.77* (0.39) | 0.5* (0.2) | -0.72 (0.39) | 0.49 (0.19) | -1.08 (1.02) | 0.3 (0.3) | -0.65 (0.44) | 0.5 (0.2) |
| Hispanic | -0.60* (0.30) | 0.5* (0.2) | -0.56 (0.3) | 0.57 (0.17) | 0.58 (0.87) | 1.8 (1.6) | -0.77* (0.33) | 0.5* (0.2) |
| Native | -15 (240) | 0 (0) | -15 (240) | 0 (0) | -15 (880) | 0 (0) | -15 (270) | 0 (0) |
| Other | -1.2* (0.5) | 0.3* (0.1) | -1.14* (0.48) | 0.32* (0.15) | -0.62 (0.95) | 0.5 (0.5) | -1.29* (0.56) | 0.3* (0.2) |
| Undisclosed | -0.25* (0.1) | 0.8* (0.1) | -0.35** (0.13) | 0.7** (0.09) | -0.21 (0.29) | 0.8 (0.2) | -0.39** (0.14) | 0.7** (0.1) |
| Female | -0.16 (0.1) | 0.9 (0.1) | -0.22 (0.13) | 0.8 (0.11) |  |  |  |  |
| non-US | 0.09 (0.1) | 0.9 (0.1) |  |  |  |  |  |  |

Validity analyses conducted on an entire population assume that factors affecting the outcome do not vary categorically or in magnitude for sub-groups. We make no such assumptions, based on anecdotal evidence in physics and published research in other disciplines that the GRE's validity may vary by subgroups [27, 28, 29]. In addition to estimating the model on the entire sample, we therefore stratified by gender and citizenship, and model PhD completion separately for these samples. We report results for the samples of *US Female*, *US Male*, *US Only*, and *All Students*. Table 2 reports regression coefficients in terms of both *logit* and odds ratio with their associated standard errors for the four different analytic samples. We summarize the findings of the analysis first by model, then by parameter.

### Findings by Model

Starting with US women, the only factors found to correlate with PhD completion are undergraduate GPA and graduate program ranking. All else in the model equal, each additional point on the GPA scale is associated with 2.5 times higher odds of PhD completion ($p = 0.02$); similarly, US women in Tier1 programs have 2.5 times higher odds of PhD completion than other programs ($p = 0.008$). Among US women, none of the GREs were found to have significant relationships with PhD completion. We find no differences by race in the probability of degree completion, though the total number of underrepresented women in the sample was small (fewer than 10 for each race).

We have similar findings on the sample of US men: undergraduate GPA and graduate program ranking are statistically significant predictors of completion. Each additional point in undergraduate GPA is associated with 1.6 times higher odds of PhD completion ($p = 0.01$); students enrolling in PhD programs ranked 55 or better have two times higher odds of completing than those in lower-ranked programs ($p < 0.001$). As with US women, no GRE has a significant relationship with PhD completion among US men in the sample. Those who identify as Hispanic ($p = 0.02$), Other ($p = 0.02$), and those for whom race data was unavailable ($p = 0.005$) have lower odds of degree completion than white students.

The results for an aggregate sample of only US students resemble those for the *US Male* sample. This similarity is to be expected since men constitute about 80% of the US sample, and thus dominate the analysis. All else in the model equal, gender is not a predictor of PhD completion. Unlike separate models of US men and women, however, GRE-Q is positively associated with PhD completion for the combined sample ($p = 0.048$), perhaps due to the larger sample size.

Estimating the model with all students in this set of PhD programs, including both US and international students, we find results similar to the *US Only* model. The statistical significance of the undergraduate GPA is reduced ($p = 0.01$) and the effect size is about halved, while that of GRE-Q increases ($p = 0.003$) but remains of similar magnitude. In this model, being in a Tier 1 program is still associated with two times higher completion odds, whereas the difference between Tier 2 and Tier 3 programs diminishes. All else in the model equal, neither gender nor citizenship status predicts PhD completion. New here, is the indication that Black students have lower odds of completion. However, considering the odds ratio and standard error increased marginally, this is likely to be a Type-I error (false positive).

### Findings by Parameter

UGPA is the only parameter that remained statistically significant across all models (Tab. 2). It has the greatest differential for PhD completion, as indicated by its positive slope in Fig. 0. Using the *US Only* model as an example, women and men with UGPAs of 4.0 have completion probabilities 14% and 12% greater than those with UGPAs of 3.0, respectively.

GRE-P scores were not associated with PhD completion at the 0.05 level of statistical significance for any of the models. This is notable because of the large span of scores among graduate students in our sample. As can be seen from Fig. 0, students scoring below the 50$^{th}$ percentile successfully complete the PhD, and they do so at a rate similar to those who scored higher. Given the limited statistical significance, the practical significance is also low: for the *US Only* model, women and men scoring at the 90$^{th}$ percentile for those groups only have completion probabilities 7% greater than those at the 10$^{th}$ percentile.

GRE-Q scores were associated with PhD completion in two models: *All Students* ($p = 0.003$), and *US Only* ($p = 0.048$). For the latter, the parameter estimate increases slightly when the US male and female populations are combined and the standard error decreases slightly, in accord with increasing the sample size. These factors, together, yield a GRE-Q parameter for the *US Only* model where $p$= 0.048. As such, and as with GRE-P, many of the highest scoring GRE-Q students do not complete, and many lower scoring students do complete. The practical significance for the *US Only* model, women and men scoring at the 90$^{th}$ percentile for those groups have completion probabilities 12% and 9% greater than those at the 10$^{th}$ percentile.

GRE-V scores were not associated with physics PhD completion in any model, and were consistently the weakest predictor among the admissions criteria in the model. As an example, the logit coefficient for GRE-V within the *All Students* model was $-0.001 \pm 0.002$ (a negative coefficient means a lower completion probability for higher scores), implying that there is no relationship between probability of PhD completion and GRE-V. This is an important finding because a myth has propagated in the physics community that GRE-V is a good predictor of completion (there is also a myth that women score higher than men on GRE-V; that, too, is false).

### Strengths and Limitations

This study has two noteworthy strengths linked to two noteworthy limitations. First, we improve on most prior GRE validity studies by including a much larger number and broader range of programs and students. However, our analysis sampled only PhD programs graduating at least 10 students per year, and findings therefore may not generalize to smaller physics programs. Thus, our sample may not generalize to the entire discipline. Our sample does represent programs with NRC rank ranging from 5 to 105, yielding a mix of more and less selective programs. Within them, the sample includes students whose GRE scores represent the tests' full dynamic range of physics major test takers, minimizing the risk of attenuation bias in our validity estimates. However, it is common for validity studies [30] to focus on Classification of Instructional Programs (CIP) Codes, whereas we focus exclusively on physics, implying that our results are likely limited to physics. Note that physics falls under CIP Code 40 [20], along with: Astronomy and Astrophysics; Atmospheric Sciences and Meteorology; Chemistry; Geological and Earth Sciences/Geosciences; and Materials Sciences.

Next, although a strength of our methodology is the use of multivariate regression to improve on the usual use of bivariate correlation as a measure of validity (i.e., by controlling for confounding factors), regression results are still correlational and our model contains omitted variable bias. As such, parameters must be interpreted in terms of association with PhD completion, with the understanding that additional factors associated with completion [31] are missing from our model. For example, program rank is positively associated with completion, but we cannot determine here the extent to which this relies on higher ranked programs selecting students with a greater overall proclivity to complete the PhD, or because higher-ranked programs have more resources with which to support students, or because students with more perseverance and drive may tend to apply to higher ranked schools. Similarly, it is unclear the extent to which lower PhD completion odds for students identifying as Black, Hispanic, and Other in some model estimations may be a function of factors that programs can control, such as the quality of advising and mentoring, willingness to accommodate a range of preparation levels, and climate for diversity. Further research is needed to understand the roles of these and other student and program-level factors to gain a full picture of PhD completion in physics, and thus identify strategies for reducing racial and gender disparities.

## Discussion

Among the parameters studied in this sample, we find PhD program rank and student undergraduate GPA are consistently associated with PhD completion, and we find consistent null results for the validity of GRE-V and GRE-P. GRE-Q has a significant relationship with PhD completion among US students as a group and all students (independent of citizenship), but not in samples of US females or US males separately.

The parameter with the strongest, consistent relationship with PhD completion is program rank, with probability of completion significantly higher for more highly ranked programs. As noted above, this study cannot pinpoint how program resources, a more stringent admissions process, and/or the self-selection of applicants to programs may explain this result. Across models, UGPA is the only admissions criterion that consistently predicts PhD completion. In weighing college grades, admissions decision makers should be cognizant that public universities, where the vast majority of underrepresented minority students earn baccalaureate degrees [32], award grades about a third of a letter grade lower than private universities [33, 34]. Thus, applying UGPA thresholds would indirectly favor White students, posing a risk to broadening participation aims.

Across models, gender, citizenship, GRE-V, and GRE-P have no bearing on PhD completion. When separately analyzing samples of US females and US males, we see no differences in PhD completion probabilities by GRE-Q, GRE-V, or GRE-P. Only when these samples are combined to increase statistical power does one of these (GRE-Q) reach conventional levels of statistical significance ($p$=0.048). The generally weak validity of GRE scores can be explained a few ways. First, factors related to the testing experience are important. Stereotype threat and test anxiety are real [7, 35], and test taking strategies can be learned by those able to afford coaching [36]. With respect to the latter, not all students receive mentoring about the importance of the tests, and thus may not undertake serious test preparation. Second, we can interpret weak GRE validity as a function of what it does and does not measure: a single exam taken on a single day represents a small sampling of student skills, not one's comprehensive capabilities, especially given that such exams are not designed to measure research potential. Third, the standardized exam format is at odds with the culture in US physics; even the subject test fails to capture how we train undergraduates and the problem solving abilities we expect of graduate students. Undergraduate physics programs in the US train students to solve complex problems that require hours or days of concerted effort. We know of no undergraduate physics programs in the US that rely on multiple choice exams in courses designed for majors. Finally, and specifically with regard to the GRE-P, the topics covered are out of phase with the typical undergraduate physics curriculum in the US. For example, large fractions of students take quantum mechanics and statistical mechanics in their senior year, either in the same semester as the GRE-P or after it has been offered. Similarly, many smaller institutions, such as minority serving institutions and liberal arts colleges, often do not have the means to offer a full suite of advanced undergraduate physics coursework. This is important to consider because about 40% of US physics undergraduates come from departments whose highest offered degree is the bachelor's [37]. The potential of students from such institutions to succeed in graduate school may thus not be represented by their GRE-P performance.

## Implications

These findings have significant implications for shaping the future of physics in the United States because the GREs are deeply entrenched in the culture of physics.

Despite compelling arguments against the use of cut off scores by the test maker itself, roughly 25% of physics PhD programs have stated minimum scores for admission on the GRE-P and GRE-Q. Perhaps more concerning are recent research findings that suggest up to 40% of US physics programs use cut off scores in practice [19]. Such decontextualized use of GRE scores embodies an admissions process that systematically filters out women of all races and national origins, Hispanics, Blacks, and Native peoples of all genders, and gives preference to international students over US students. The weight of evidence in this paper contradicts conventional wisdom, and indicates clearly that lower than average scores on admissions exams do not imply a lower than average probability of earning a physics PhD. Continued overreliance on metrics that do not predict PhD completion but have large gaps based on demographics works against both the fairness of admissions practices and the health of physics as a discipline.

This study implies an urgent need for additional research that improves admissions processes in physics and beyond. The community ought to reevaluate admissions criteria and practices to ensure that selection is both equitable and effective for identifying students that can be successful. This effort will require identifying both a broader set of applicant characteristics that predict graduate student outcomes, as well as understanding the characteristics of mentoring and PhD programs that create healthy learning environments. For example, our finding that program ranking was the strongest and most consistent single predictor tells us that the context in which doctoral students are admitted and learn matters for their success.

Further, following the assumption that GRE-P signals preparation in the discipline, the subject test's limited validity in predicting completion implies that disciplinary preparation itself may be necessary, but insufficient, to identify completers. Discriminating on GRE implied preparation, rather than a holistic assessment of potential to earn the PhD and conduct research, overlooks applicants who could become strong research physicists. Excellence as a researcher is likely also a function of research mentoring and experience (both before and in graduate school) and socio-emotional/non-cognitive competencies (e.g., initiative, conscientiousness, accurate self-assessment, communication), which scholars have linked to performance in other professional and educational domains [38]. It is time to think creatively about both assessing such qualities alongside academic preparation as part of a holistic approach to graduate admissions, as well as strategies that connect prospective students to graduate programs in which they will thrive.

## Materials and Methods

The IPEDS data in Tab. 0 spanned the years 2009 through 2014; the 5-year average is reported. To count as a physics degree, we added all degree classifications that are primarily given in a physics department. These numbers agree well with those independently gathered by the Statistical Research Center of the American Institute of Physics. The number of first year graduate students was obtained from the National Science Foundation (NSF) and National Institutes of Health (NIH) Survey of Graduate Students and Postdoctorates in Science and Engineering [20]. IPEDS does not separate MS from PhD students; we estimate that 90% of the incoming graduate students are intending to pursue the PhD [39].

The validity of various factors in predicting PhD completion may vary by student demographic characteristics, national origin, and program selectivity. For example, considering that the GRE was originally developed on samples of men, we might anticipate that GRE-Q, GRE-V, and/or GRE-P would, individually, have stronger relationships with PhD completion in samples of men than in samples of women. Similarly, cultural differences around frequency of standardized testing may advantage students from countries where such practices permeate the higher education system more than others. Therefore, we stratified the sample by gender, citizenship, and ranking tier, and conduct the multivariate regression on each. Although narrowing the analytic sample with this approach reduces statistical power, the sample sizes here are more than sufficient for the employed regression methods to detect a reasonably sized effect. Multiple imputation (multivariate normal algorithm) was necessary to impute missing data for undergraduate GPA and GRE-P. This had minimal effect on the results, but increased the sample size by about twenty percent.

Given that the median time to degree across physics PhD programs is 6 years, some students who started before 2010 were still active at the time of data collection in 2016. The probability of not completing the physics PhD has an exponential time dependence with a time constant of 1.8 years. Thus, students who have been in their programs for three time constants have only a 5% chance of not completing. These students were thus categorized as completers in this study.

ETS changed its general test scale from 200-800 to 130-170 in 2009. Thus, scores were converted to percentiles for each test using concordance tables from ETS to obtain comparable measures across the study.

We conducted a variety of sensitivity tests to ensure the reliability of our findings. Most importantly, we replicated the analyses in both Stata/SE 14.2 and R 3.3.3. Coefficients and p-values were equivalent to at least the tenths place in all but a handful of cases, which could be explained by different samples generated by the different multiple imputation packages.

To reduce bias in our estimates that could come from year-to-year variation, we included start year fixed effects. We also examined the stability of coefficients to a variety of model specifications. Given the large share of missing race data, for example, we separately included a dichotomous variable for *Race Unavailable* and used only cases for which race data was available. Results via these approaches did not vary substantively, so we used the former to maximize the analytic sample.

Our goal here was not to identify the best predictive model with the minimum number of parameters, but rather to understand how all four commonly used admissions metrics (GPA, GRE-Q, GRE-V, and GPA) and the most salient demographic information would contribute to a discussion of metrics and diversity by admissions committees. That said, we did conduct sensitivity tests that examined tested bivariate relationships and added variables to the model stepwise, to ensure we capture both individual relationships and well as how they operate together to explain PhD completion. However, we report only selected models for the sake of parsimony.

Finally, we investigated program-based weights to understand if variations in the number of records from individual programs would affect our estimates. Weighting schemes had negligible effect on the results, and were thus not used in the analyses reported here.

**Acknowledgements:** The authors thank J. Pelz and H. Lewandowski for useful comments. **Funding:** CWM was supported initially by NSF-CAREER 1522927 and finally by NSF 1633275; BMZ was supported by NSF 1633275; JRP was supported by NSF-INCLUDES 1649297; TH was supported by NSF 1143070. **Author Contributions:** CWM and TH designed the research; CWM, BZ, JRP conducted the research; BZ, JRP, and RTS performed statistical analyses; CWM, BZ, JRP, and TH discussed results and manuscript, and wrote the paper. **Competing interests:** The authors declare that they have no competing interests. **Data and materials availability:** All data needed to evaluate the conclusions of the paper are present in the paper. Additional data related to this paper may be requested from the authors. **IRB:** The University of Maryland College Park Institutional Review Board determined this project to be EXEMPT FROM IRB REVIEW according to federal regulations.

# Appendices

The goal of this section is to address questions and comments regarding our article.

## Correlations and Collinearity

One of our initial steps in assessing the data set collected for this work was to perform a principal components analysis to estimate the potential impact of any collinearity among the four input metrics (ug.GPA, GRE.V, GRE.Q, GRE.P). The results of this analysis indicated that the correlations between the variables of interest were not significant enough to cause concern for including them as independent variables in a multiparameter regression. This conclusion is further borne out in the correlation matrices and the variance inflation factors, as discussed below.

The bivariate correlations among the four continuous variables in each of the four analytic samples ("All students", "US only", "US female", and "US male") are included in Appendix Ia, and the correlations among the resultant fit parameters are included in Appendix Ib. The correlations are all weak to moderate. While it is understandable to presume that the correlation between GRE.Q and GRE.P might be very strong, the data do not substantiate that assumption. Some possible reasons for lower than expected correlation between these includes: (a) the GRE.Q only tests elementary mathematical skills ("...high school mathematics and statistics at a level that is generally no higher than a second course in algebra; it does not include trigonometry, calculus or other higher-level mathematics.") This leads to significant clustering of GRE.Q scores at the upper end of the possible scores for test takers that are physics majors; (b) the GRE.P is administered early in the fall and late in the spring, meaning that many US students are either concurrently learning or have not yet begun formal coursework addressing all of the advanced physics topics that the GRE.P includes.

Multicollinearity can also be measured by a variance inflation factor (VIF). When two variables are independent (orthogonal, zero correlation), the VIF is 1. As the correlation among variables increases, so will the VIF values. Statisticians agree that when VIF values are between 1 and 5, there is not much concern including the correlated variables; when the VIF values exceed 10, multicollinearity is an issue that might lead to errors in interpretation. More information about this can be found in Silvestrini and Burke (2018) or O'Brien (2007). As reported in Appendix II, the VIF calculated for each analytic sample was 2.0 or below. The results were similar between imputations, and are thus reported for just one imputation. Neither the correlation matrices nor VIF analyses raise sufficient concerns about multicollinearity to deter multivariate regression with all four continuous variables. A multivariate approach is desirable because it enables more precise estimates of individual variables' associations with the outcome, and recent evidence that admissions in most physics PhD programs rely on all of these criteria.

## Use of Program Tier as a Categorical Variable

The goal of our analysis was predicting how GRE scores and GPA may be associated with PhD completion. We accounted for tier of the PhD program to create more precise estimates of these relationships, reduce omitted variable bias in explaining PhD completion overall, and to address selection bias.  Post-stratification (e.g., the use of tier to average across programs of similar ranking)   is a common method used to deal with selection bias.

Additionally, this approach allows us to speak to multiple audiences.  For example, it is common for elite programs to disregard analyses when the sample includes non-elite programs, and vice versa.  Our choice to stratify the sample allows different audiences to understand how students fare at peer or near-peer institutions.

# Relative Quality of Models

As we noted in the article, our goal was not to determine which if all possible models could best explain the outcome of interest.  We reported on only four of many models that we tested.   We did this because (a) these models include all the variables available to and used by admissions committees, and (b) alternate models did not lead to major differences in the findings.  To support the latter claim, and to provide additional perspective about the appropriateness of models reported in the paper, Appendix III shows that the models we reported on are comparable to or superior to several alternate models, as indicated by multiple metrics, namely the Akaike Information Criteria (corrected for sample size, AICc) and the Bayesian Information Criteria (BIC). For these metrics, differences of two or less indicate models of equivalent quality; a model whose metric is lower by 2-6 than another model's metric is the better of the two; models whose metrics are lower by six or more are substantially better models.

A few alternate models reported on in the present correspondence include: (a) excluding GRE.Q from the four original parameters; (b) excluding GRE.P from the four original parameters; (c) including only GRE.Q; (d) including only GRE.P; (e) using the average of GRE.Q and GRE.P along with ug.GPA and GRE.V; and (f) including only ug.GPA.

According to the AICc, all four of the models included in our article are equivalent or superior to each of these alternate models above with one exception: including only ug.GPA (alternate model (f)) is a better model than ours for the sample of US Women.  Furthermore, our models were the best-fitting models, or equivalent to the best-fitting model, for each of the samples, again with the exception of US Women, for which ug.GPA was the best-fitting model.

Using the BIC, the three best-fitting models via are all single-parameter models.    For three of the samples—US Only, US Female, and US Male—the best-fi model includes only ug.GPA.  For the All Students model, including only GRE.Q was the best-fitting model.   The BIC's penalty for additional parameters disadvantages multiparameter models like the ones we used to model the current admissions process.  While using undergraduate GPA alone may be somewhat reasonable, given that this averages across many different courses, topics, and years, it would be difficult to seriously suggest that programs use only the GRE.Q--which again, tests math at about the 8th or 9th grade level--and no other common information to select who gets into PhD programs simply because it scored well through the BIC.

### p-values

We used a p-value of 0.05 to indicate statistical significance, with a note indicating that issues with relying on p-value are known (e.g., risk of observing Type I errors-false positives-increases as sample size increases)    The nebulous nature of the p-value is in part why we commented throughout the article on the practical significance of the various parameters, i.e., reporting both a model's p-value and probabilities predicted by the model for a range of inputs.   The likelihood of finding statistical significance at a specific level is greater with larger sample sizes.    It is worth noting that our sample sizes are large relative to analyses conducted by ETS to validate the GRE tests: a recent validity study used 508 students in CIP Code 40 to validate the GRE for doctoral students in the physical sciences.  Our analytic samples "All Students", "US only", and "US male" contain 4-6 times as many data, making the likelihood of finding a statistically significant result higher.  Our "US female" sample had N = 402, which is comparable to what was used in ETS studies (though in our study, those are all physics students, not a mixture of students from the disciplines comprising CIP code 40).

The concept of p-hacking implies that researchers search, even if subconsciously, for models and subsets of the data that allow them to find and report a preferred conclusion.  As noted above, the models reported on had strong theoretical rationales for their use, were as good as alternate models, and subsetting of the data was done with very broad categories that have a sound theoretical foundation:  ETS data, and long standing observations within the physics community, reveal that men outscore women, and international students outscore US citizens on GRE tests.   Therefore, subsetting the analytic sample by gender and citizenship status is reasonable.

## Restriction of Range

All validity studies of this sort, including those conducted by test makers to validate their products, suffer from restriction of range. That is, validity estimates are likely to be attenuated by a narrower range of scores earned by people who were accepted into PhD programs, and thus included in any such study, as compared to the typically larger range of scores of all test takers, including those not admitted. The simple fact is that the academic performance of students who do not matriculate cannot be studied.  Having said that, restriction of range is not as significant of an issue in our study as is commonly assumed for works of this nature because we have the testing data for the subset of test takers intending to go to graduate school for physics.   As the table in Appendix IV shows, the ranges of our analytic sample relative to the group of physicist test takers are comparable for the physics and verbal GRE tests.    Given the similarity in range for most of the factors, corrections for restriction of range are not likely to significantly alter our findings for GRE.P and GRE.V.  The range of GRE.Q is restricted in our sample to about half of the range for all physicist test takers.   Corrections with respect to GRE.Q are likely to yield somewhat stronger correlations.   Note that physicists' scores are significantly restricted relative to the overall test taker population; because of that, the subset of test takers that are physicists is the appropriate group to compare to our analytic samples for restriction of range.

# Appendix I: Correlation Matrices

## a: Correlations among independent parameters

The following matrices show the correlations between the four quantitative admission metrics within the dataset. Correlation coefficients are Spearman's rank order correlation averaged across 40 imputations of missing data.

|  |  | ug.GPA | GRE.Q. | GRE.V. | GRE.P. |
|---|---|---|---|---|---|
| **All Students** | ug.GPA | 1.00 | 0.18 | 0.18 | 0.18 |
|  | GRE.Q. | 0.18 | 1.00 | 0.23 | 0.55 |
|  | GRE.V. | 0.18 | 0.23 | 1.00 | 0.14 |
|  | GRE.P. | 0.18 | 0.55 | 0.14 | 1.00 |
| **US Only** | ug.GPA | 1.00 | 0.26 | 0.18 | 0.28 |
|  | GRE.Q. | 0.26 | 1.00 | 0.35 | 0.55 |
|  | GRE.V. | 0.18 | 0.35 | 1.00 | 0.33 |
|  | GRE.P. | 0.28 | 0.55 | 0.33 | 1.00 |
| **US Male** | ug.GPA | 1.00 | 0.24 | 0.16 | 0.36 |
|  | GRE.Q. | 0.24 | 1.00 | 0.34 | 0.54 |
|  | GRE.V. | 0.16 | 0.34 | 1.00 | 0.34 |
|  | GRE.P. | 0.36 | 0.54 | 0.34 | 1.00 |
| **US Female** | ug.GPA | 1.00 | 0.26 | 0.22 | 0.24 |
|  | GRE.Q. | 0.26 | 1.00 | 0.45 | 0.61 |
|  | GRE.V. | 0.22 | 0.45 | 1.00 | 0.42 |
|  | GRE.P. | 0.24 | 0.61 | 0.42 | 1.00 |

## b: Correlations among fit parameters

The following matrix showing correlations among the fit parameters resulting from the indicated models was averaged over 40 imputations. Only the 4 quantitative metrics are shown. Demographic factors (race, citizenship, gender), tier, and start.year fixed effect were also calculated, but are not shown here.

| | | (Intercept) | ug.GPA | GRE.Q | GRE.V | GRE.P |
|---|---|---|---|---|---|---|
| **All Students** | (Intercept) | 1.00 | -0.71 | -0.49 | -0.03 | 0.22 |
| | ug.GPA | -0.71 | 1.00 | -0.14 | -0.08 | -0.06 |
| | GRE.Q | -0.49 | -0.14 | 1.00 | -0.16 | -0.42 |
| | GRE.V | -0.03 | -0.08 | -0.16 | 1.00 | -0.17 |
| | GRE.P | 0.22 | -0.06 | -0.42 | -0.17 | 1.00 |
| | | (Intercept) | ug.GPA | GRE.Q | GRE.V | GRE.P |
| **US Only** | (Intercept) | 1.00 | -0.74 | -0.37 | -0.10 | 0.18 |
| | ug.GPA | -0.74 | 1.00 | -0.18 | -0.06 | -0.04 |
| | GRE.Q | -0.37 | -0.18 | 1.00 | -0.20 | -0.42 |
| | GRE.V | -0.10 | -0.06 | -0.20 | 1.00 | -0.13 |
| | GRE.P | 0.18 | -0.04 | -0.42 | -0.13 | 1.00 |
| | | (Intercept) | ug.GPA | GRE.Q | GRE.V | GRE.P |
| **US Male** | (Intercept) | 1.00 | -0.74 | -0.38 | -0.12 | 0.20 |
| | ug.GPA | -0.74 | 1.00 | -0.17 | -0.05 | -0.06 |
| | GRE.Q | -0.38 | -0.17 | 1.00 | -0.17 | -0.41 |
| | GRE.V | -0.12 | -0.05 | -0.17 | 1.00 | -0.15 |
| | GRE.P | 0.20 | -0.06 | -0.41 | -0.15 | 1.00 |
| | | (Intercept) | ug.GPA | GRE.Q | GRE.V | GRE.P |
| **US Female** | (Intercept) | 1.00 | -0.75 | -0.35 | -0.04 | 0.15 |
| | ug.GPA | -0.75 | 1.00 | -0.19 | -0.08 | 0.04 |
| | GRE.Q | -0.35 | -0.19 | 1.00 | -0.27 | -0.43 |
| | GRE.V | -0.04 | -0.08 | -0.27 | 1.00 | -0.06 |
| | GRE.P | 0.15 | 0.04 | -0.43 | -0.06 | 1.00 |

# Appendix II: Variance Inflation Factors

The variance inflation factor was computed using the `vif()` function within the `car` package in R version 3.5.1. The standard Variance Inflation Factor (VIF) is computed when each variable represents just one degree of freedom (e.g., ug.GPA). However, the Generalized Variance Inflation Factor (GVIF) is automatically used when models include some categorical variables, such as race or tier, which have multiple discrete values (e.g., Tier 1, Tier 2, Tier 3). Generalized Variance Inflation Factor computed for the models gin Table 2 of our Science Advances paper are below.

| | Parameter | GVIF | Df | GVIF^(1/(2*Df)) |
|---|---|---|---|---|
| All Students | ug.GPA | 1.14 | 1 | 1.07 |
| | GRE.Q. | 1.56 | 1 | 1.25 |
| | GRE.V. | 1.30 | 1 | 1.14 |
| | GRE.P. | 2.02 | 1 | 1.42 |
| | gender | 1.07 | 1 | 1.03 |
| | tier | 1.30 | 2 | 1.07 |
| | race | 1.93 | 7 | 1.05 |
| | start.yr | 1.11 | 10 | 1.01 |
| US Only | ug.GPA | 1.16 | 1 | 1.08 |
| | GRE.Q. | 1.56 | 1 | 1.25 |
| | GRE.V. | 1.20 | 1 | 1.09 |
| | GRE.P. | 1.71 | 1 | 1.31 |
| | gender | 1.09 | 1 | 1.04 |
| | tier | 1.55 | 2 | 1.12 |
| | race | 1.38 | 6 | 1.03 |
| | start.yr | 1.12 | 10 | 1.01 |
| US Male | ug.GPA | 1.17 | 1 | 1.08 |
| | GRE.Q. | 1.51 | 1 | 1.23 |
| | GRE.V. | 1.18 | 1 | 1.09 |
| | GRE.P. | 1.65 | 1 | 1.28 |
| | tier | 1.60 | 2 | 1.12 |
| | race | 1.40 | 6 | 1.03 |
| | start.yr | 1.13 | 10 | 1.01 |
| US Female | ug.GPA | 1.21 | 1 | 1.10 |
| | GRE.Q. | 1.81 | 1 | 1.35 |
| | GRE.V. | 1.31 | 1 | 1.15 |
| | GRE.P. | 1.71 | 1 | 1.31 |
| | tier | 1.55 | 2 | 1.12 |
| | race | 1.58 | 6 | 1.04 |
| | start.yr | 1.37 | 10 | 1.02 |

# Appendix III: Relative Quality of Models

AICc and BIC values for various models. Dark shading (yellow for AIC, blue for BIC) indicates the minimum value. Models that are 2 or less from the minimum, signaling models equivalent to the minimum, have lighter shading.

| Model | Continuous variables included | Continuous variables excluded | All Students AICc | All Students BIC | US Only AICc | US Only BIC | US Male AICc | US Male BIC | US Female AICc | US Female BIC |
|---|---|---|---|---|---|---|---|---|---|---|
| Sci. Adv. | ug.GPA GRE.Q GRE.V GRE.P | | 4159 | 4315 | 2435 | 2572 | 1986 | 2114 | 484 | 573 |
| a | ug.GPA GRE.V GRE.P | GRE.Q | 4163 | 4313 | 2435 | 2566 | 1986 | 2108 | 482 | 568 |
| b | ug.GPA GRE.Q GRE.V | GRE.P | 4159 | 4309 | 2435 | 2567 | 1987 | 2108 | 482 | 567 |
| c | GRE.Q | ug.GPA GRE.V GRE.P | 4161 | 4299 | 2443 | 2563 | 1989 | 2100 | 483 | 561 |
| d | GRE.P | ug.GPA GRE.Q GRE.V | 4166 | 4305 | 2444 | 2565 | 1990 | 2101 | 485 | 562 |
| e | ugGPA (GRE.P+GRE.Q)/2 GRE.V | | 4159 | 4309 | 2433 | 2565 | 1985 | 2107 | 482 | 567 |
| f | ug.GPA | GRE.Q GRE.V GRE.P | 4166 | 4304 | 2437 | 2557 | 1986 | 2097 | 479 | 557 |

# Appendix IV: Restriction of Range

The table below reports the interquartile ranges (IQR) for groups of GRE test takers and our analytic sample. Additionally included is the range 75th-25th; comparison of these ranges between the test taker distributions (lightly shaded rows) and our sample (unshaded rows) for each group gives an idea of range restriction. Rows with dark shading indicate the IQR for the related overall test taker population, independent of intended graduate major. The following nomenclature is used below: "physicists" = individuals whose self-identified intended graduate major is physics, as indicated in ETS database; "US" = individuals identifying as US citizens, as indicated in ETS database; "male"= individuals identifying as male, as indicated in ETS database; "female"= individuals identifying as female, as indicated in ETS database.

| Test | Group | 25th | 50th | 75th | Range 75th-25th |
|---|---|---|---|---|---|
| GRE.P | Physicists | 26 | 51 | 75 | 49 |
| | Our Sample: physicists | 42 | 65 | 85 | 43 |
| | US physicists | 28 | 40 | 63 | 35 |
| | Our Sample: US physicists | 35 | 55 | 71 | 36 |
| | US male physicists | 21 | 42 | 67 | 46 |
| | Our Sample: US male physicists | 39 | 57 | 73 | 34 |
| | US female physicists | 13 | 26 | 47 | 34 |
| | Our Sample: US female physicists | 21 | 37 | 57 | 36 |
| GRE.Q | Physicists | 69 | 81 | 91 | 22 |
| | Our Sample: physicists | 81 | 89 | 91 | 10 |
| | US | 20 | 38 | 59 | |
| | US physicists | 62 | 78 | 87 | 25 |
| | Our Sample: US physicists | 79 | 87 | 91 | 12 |
| | US male | 27 | 51 | 69 | |
| | US male physicists | 66 | 78 | 89 | 23 |
| | Our Sample: US male physicists | 79 | 87 | 91 | 12 |
| | US female | 17 | 30 | 51 | |
| | US female physicists | 59 | 73 | 84 | 25 |
| | Our Sample: US female physicists | 75 | 83 | 91 | 16 |
| GRE.V | Physicists | 35 | 65 | 86 | 51 |
| | Our Sample: physicists | 57 | 77 | 89 | 32 |
| | US | 39 | 61 | 80 | |
| | US physicists | 65 | 83 | 93 | 28 |
| | Our Sample: US physicists | 68 | 81 | 91 | 23 |
| | US male | 43 | 69 | 86 | |
| | US male physicists | 65 | 80 | 93 | 28 |
| | Our Sample: US male physicists | 68 | 81 | 91 | 23 |
| | US female | 35 | 56 | 76 | |
| | US female physicists | 69 | 83 | 93 | 24 |
| | Our Sample: US female physicists | 70 | 83 | 93 | 23 |